\documentclass[twoside,twocolumn]{article}

\usepackage{blindtext} 
\usepackage{subcaption}

\usepackage{graphicx}
\usepackage[sc]{mathpazo} 
\usepackage[T1]{fontenc} 
\usepackage{microtype} 
\usepackage{flushend}
\usepackage[english]{babel} 
\usepackage[hmarginratio=1:1,top=32mm,columnsep=20pt]{geometry} 
\usepackage{caption} 
\usepackage{booktabs} 
\usepackage{float}
\usepackage{placeins}
\usepackage[super,numbers]{natbib}
\usepackage{balance}

\usepackage{enumitem} 
\setlist[itemize]{noitemsep} 

\usepackage{glossaries}
\setacronymstyle{long-short}
\newacronym{pes}{PES}{potential energy surface}
\newacronym{mep}{MEP}{minimal energy path}
\newacronym{neb}{NEB}{Nudged Elastic Band}
\newacronym{cineb}{CINEB}{Climbing Image Nudged Elastic Band}
\newacronym{gsm}{GSM}{Growing String Method}
\newacronym{idpp}{IDPP}{Image Dependent Pair Potential}
\newacronym{dft}{DFT}{Density Functional Theory}
\newacronym[plural=Neural Networks]{nn}{NN}{Neural Network}
\newacronym{mpnn}{MPNN}{Message Passing Neural Network}
\newacronym{mse}{MSE}{Mean Squared Error}
\newacronym{ci}{CI}{Climbing Image}
\newacronym{ml}{ML}{Machine Learning}
\newacronym{ase}{ASE}{Atomic Simulation Environment}
\newacronym{hdf5}{HDF5}{Hierarchical Data Format}
\newacronym{mae}{MAE}{Mean Average Error}
\newacronym{rmse}{RMSE}{Root Mean Square Error}
\newacronym{qm}{QM}{quantum mechanical}
\newacronym{md}{MD}{molecular dynamics}
\newacronym{qbc}{QbC}{Query by Comittee}
\newacronym{ani1x}{ANI1x}{ANI1x \cite{ani1x}}

\usepackage{titlesec} 
\renewcommand\thesection{\Roman{section}} 
\renewcommand\thesubsection{\roman{subsection}} 

\titleformat{\section}[block]{\large\scshape\centering}{\thesection.}{1em}{} 
\titleformat{\subsection}[block]{\large}{\thesubsection.}{1em}{} 

\usepackage{fancyhdr} 
\usepackage{titling} 
\usepackage{hyperref} 

\title{Transition1x - a Dataset for Building Generalizable Reactive Machine Learning Potentials \\ \vspace{-10pt} \hrulefill \vspace{-28pt}} 
\author{%
Mathias Schreiner\textsuperscript{\rm 1}, Arghya Bhowmik\textsuperscript{\rm 2}, Tejs Vegge\textsuperscript{\rm 2}, Jonas Busk\textsuperscript{\rm 2}, Ole Winther\textsuperscript{\rm 1,3,4} \vspace{4pt}
\\  
\textsuperscript{\rm 1} \normalsize DTU Compute, Technical University of Denmark (DTU)\\
\textsuperscript{\rm 2} \normalsize DTU Energy, Technical University of Denmark (DTU)\\
\textsuperscript{\rm 3} \normalsize Department of Biology, University of Copenhagen (UCph)\\
\textsuperscript{\rm 4} \normalsize Genomic Medicine, Copenhagen University Hospital, Rigshospitalet\\
}
\date{} 

\renewcommand{\maketitlehookd}{%
\begin{abstract}
\noindent
\textit{
\gls{ml} models have, in contrast to their usefulness in molecular dynamics studies, had limited success as surrogate potentials for reaction barrier search. This is primarily because available datasets for training \gls{ml} models on small molecular systems almost exclusively contain configurations at or near equilibrium. In this work, we present the dataset Transition1x containing 9.6 million \gls{dft} calculations of forces and energies of molecular configurations on and around reaction pathways at the $\omega$B97x/6-31G(d) level of theory. The data was generated by running \gls{neb} with \gls{dft} on 10k organic reactions of various types while saving intermediate calculations. We train equivariant graph message-passing neural network models on Transition1x and cross-validate on the popular ANI1x and QM9 datasets. We show that \gls{ml} models cannot learn features in transition state regions solely by training on hitherto popular benchmark datasets. Transition1x is a new challenging benchmark that will provide an important step towards developing next-generation \gls{ml} force fields that also work far away from equilibrium configurations and reactive systems. }
\end{abstract}
}

\begin{document}

\maketitle
\section{Introduction}
\gls{ml} models for molecular systems have accuracy comparable to \gls{qm} methods but the computational cost of classical interatomic potentials\cite{Faber2017NNDFT,Westermayr2021ml4qm1,Campbell2020ml4qm2,behler2007generalized,Westermayr2021qm4ml4}. The development of such data-driven models has ushered in a new age in computational chemistry over the last few years\cite{unke2021machine, behler2021four, huang2021ab,deringer2021gaussian}. \gls{ml} potentials have been used for a variety of tasks such as structural optimization \cite{Kaappa2021structuraloptimization} or the study of finite-temperature dynamical properties through molecular dynamics \cite{Wang2020finitetemp}. \gls{ml} potentials are especially suited for screening through large numbers of molecules or simulating systems that are too large for traditional \gls{qm} methods due to a complexity scaling that is orders of magnitudes lower. The applicability of these models depends on the sampling of training data of chemical and structural space\cite{anatol2020chemspace}. Fitting \gls{ml} models to the entire \gls{pes} requires lots of carefully selected data as the underlying electronic interaction between atoms is of a complex, quantum mechanical nature. Thus the focus remains on an efficient sampling strategy of the useful parts of the \gls{pes} that are relevant to the application at hand. For example, models for optimization tasks should be trained on datasets including small perturbations to equilibrium geometries, and models for \gls{md} simulations and reactive systems should be trained on datasets with high energy geometries and states that represent the making and breaking of bonds. \\

\gls{ml} potentials that allow accurate modeling of general reaction barriers are challenging to train and only limited demonstrations have been shown to date. Acceptable accuracy has been achieved by focusing on single or few types of reactions involving small molecules with tractable dataset size \cite{lu2018rate, young2021transferable, manzhos2020neural, qmrxn} or by studying simple molecular dissociation \cite{malshe2007theoretical}. \gls{ml} models that can accurately predict \glspl{pes} for unseen chemical reactions must be incredibly expressive and have access to training data that extensively samples structures from reactive and high-energy regions (compared to near-equilibrium geometries) of chemical space. Recently, the development of \gls{nn} architectures that learn representation and energy/force mapping\cite{unke2021machine} has tackled the problem of expressive models, but creating datasets with millions of data points sampled around reactions of various types allowing such models to generalize across a large number of reactions has remained an open challenge. Thus, \gls{ml} potentials have not yet proved capable of accurate and general prediction of reaction barriers and transition states. 
\\

Sampling of rare transition events is efficiently done with the \gls{neb} method\cite{neb}. Here we propose a new dataset for building \gls{ml} models capable of generalizing across a large variety of reaction \glspl{pes}. We base our work on a dataset of reaction-product pairs from Grambow et al. 2020\cite{rxns}. The original dataset contains a wide range of organic reactions representing bond changes between all possible combinations of H, C, N, and O atoms. We leverage \gls{neb}-based \gls{pes} exploration as an efficient data collection tool and prove its superiority compared to \gls{md}-based dataset preparation on reactive molecular configurations by testing the accuracy of \gls{ml} models built from both types of data. Moreover, Transition1x is compatible with ANI1x in the level of \gls{dft} such that \gls{ml} models can be trained on the two in conjunction to leverage both of their strengths.
\\

Ultra-fast prediction of chemical reaction kinetics, especially for computational modeling of complex reaction networks, is groundbreaking for the entire field of chemical and molecular sciences. We believe that the Transition1x dataset will expedite the development and testing of universal reactive \gls{ml} potentials that help the community achieve that goal.

\section*{Methods}

\begin{figure*}
\centering
\includegraphics[width=\textwidth]{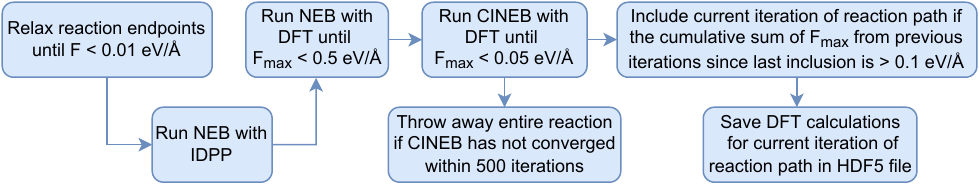}
\caption{Overview of the data generation workflow. First, reactant and product are relaxed before generating an initial \gls{mep} guess with \gls{idpp}\cite{idpp}. Next \gls{neb}\cite{neb} and \gls{cineb}\cite{cineb} is run on the initial path until convergence. If the \gls{mep} does not converge within 500 iterations we discard the reaction, as unphysical configurations may have been encountered. If the reaction converges, all intermediate paths are saved in the dataset, as long as they are sufficiently different from previously saved paths.}
\label{fig:workflow}
\end{figure*}

Starting from a set of 11961 reactions \cite{rxns} with reactants, transition states, and products, \gls{neb} is used to explore millions of molecular configurations in transition state regions, using \gls{dft} to evaluate forces and energies. The resulting \gls{dft} calculations are available in the Transition1x dataset. Figure \ref{fig:workflow} presents an overview of the workflow. Reactant and product are relaxed for any particular reaction before generating an initial path using \gls{idpp} \cite{idpp}. Next, the \gls{mep} is optimized with \gls{neb}\cite{neb} and consecutively \gls{cineb}\cite{cineb} until convergence. If the path converges, we save the \gls{dft} calculations from the iterations for which the current reaction path moved significantly.

\subsubsection*{Initial Data}
The data generating procedure starts by taking an exhaustive database \cite{rxns} of product-reactant pairs based on the GDB7 dataset \cite{gdb7}. Each reaction consists of up to seven heavy atoms including C, N, and O. The authors of this data used the \gls{gsm}\cite{gsm} with the $\omega$B97X-D3\cite{wb97x}/def2-TZVP level of theory to generate reactants, products, and transition states for 11961 reactions using Qchem \cite{qchem}.

\subsubsection*{Density Functional Theory}
For compatibility with ANI1x \cite{ani1x}, the $\omega$B97x\cite{wb97x} and 6-31G(d)\cite{6-31G} basis set is applied to perform all calculations in ORCA 5.0.2 \cite{orca}. 

\subsubsection*{Optimizer}
The BFGS optimizer \cite{bfgs} implemented in \gls{ase}\cite{ase} with $\alpha=70$ and a maximal step size of 0.03 Å is used for all optimization tasks, including relaxing endpoints and running both \gls{neb} and \gls{cineb}.

\subsubsection*{Initial Path Generation}
Product and reactant geometries are relaxed in the potential before running \gls{neb}. The configuration is considered relaxed once the norm of the forces in configurational space is less than a threshold of 0.01 eVÅ$^{-1}$. After relaxing the endpoints an initial path is proposed, built from two segments -- one interpolated from the reactant to the transition state from the original data, and another interpolated from the transition state to the product. Next, the initial path is minimised with \gls{neb} using \gls{idpp}\cite{idpp}, a potential specifically designed to generate physically realistic reaction paths for \gls{neb} at a low computational cost. Finally, the path is proposed as the initial \gls{mep} in the \gls{dft} potential. 

\subsubsection*{Nudged Elastic Band}
\gls{neb}\cite{neb} is a double ended search method for finding \glspl{mep} connecting reactant and product states. It works by iteratively improving an initial guess for the \gls{mep} by using information about the \gls{pes} as calculated by some potential. \gls{neb} represents the path as a series of configurations called images connected with an artificial spring force. The energy of the path is minimized by iteratively nudging it in the direction of the force perpendicular to it until convergence. After the path has converged, there is no guarantee that the maximal energy image represents the correct transition state as the maximal energy image may not correspond with the true maximum along the path. \gls{cineb}\cite{cineb} is an improvement to the \gls{neb} algorithm as it imposes, as an additional condition to the convergence, that the maximal energy image has to lie at a maximum. It does so by, in each iteration, letting the image with the maximal energy climb freely along the reaction path. Running \gls{cineb} from the beginning, however, can interfere with the optimizer and result in slow (or wrong) convergence of the \gls{mep} as the climbing image can pull the current path off the target \gls{mep} if the paths are not close. Therefore, first, the path is relaxed with regular \gls{neb} until the maximal perpendicular force to the \gls{mep} is below a threshold of 0.5 eVÅ$^{-1}$. At this point, \gls{neb} has usually found the qualitatively correct energy valley, and further optimization only nudges the path slightly while finding the bottom. At this point, \gls{cineb} is turned on to let the highest energy image climb along the path until it finds an energy maximum. \gls{cineb} is run until the path has been relaxed fully and the maximal perpendicular force on the path does not exceed 0.05 eVÅ$^{-1}$. This threshold was chosen as a compromise between having accurate reaction paths in the dataset and limiting redundant \gls{dft} calculations. No further refinement of the transition states was done at this point as the goal is to generate a dataset of molecular configurations close to reaction pathways rather than finding accurate transition states. Ten images were used to represent all reaction paths and the spring constant between them was 0.1 eVÅ$^{-2}$. 

\subsubsection*{Data selection}
When running \gls{neb}, unphysical configurations are often encountered in reactions that do not converge. Such images in the data will interfere with model when training, and therefore those reactions are discarded entirely. There are 10073 converged reactions in Transition1x. In the final steps of \gls{neb}, the molecular geometries of images are similar between each iteration as the images are nudged only slightly close to convergence. Data points should be spread out so that models do not overfit to specific regions of the data. Updated paths are only included in the dataset if they are significantly different from previous ones. The maximal perpendicular force, $F_{max}$, to the path is a proxy for how much the path moves between iterations. Once the cumulative sum of $F_{max}$ from previous iterations, since the last included path, exceeds 0.1 eVÅ$^{-1}$ the current path is included. This means that often in the first iterations of \gls{neb} every path is included, but as we move towards convergence new data points are included at a lower frequency. 

\subsubsection*{Model and Training}
To validate the dataset, we train and evaluate PaiNN \cite{painn} models on Transition1x, QM9x and ANI1x and compare the their performances. PaiNN is an equivariant \gls{mpnn}\cite{mpnn} model specifically designed to predict properties of molecules and materials. Forces are calculated as the negative gradient of the energy wrt. the Cartesian coordinates of the atoms rather than as a direct output from the model. This ensures consistent forces. We used a cut-off distance of 5 Å to generate the molecular graph neighborhood, three message-passing steps, and 256 neurons in each hidden layer of the model. The model was trained using the ADAM \cite{adam} optimizer and an initial learning rate of $10^{-4}$. During training, the learning rate was decreased by 20\% if no improvement was seen on training data for $10^4$ batches. The loss is a combination of a squared error loss on force and energy. The force error is the Euclidian distance between the predicted and the true force vector divided by the number of atoms in the molecule, as otherwise, the force term would contribute more to the loss on bigger molecules. All datasets are stratified by molecular formula such that no two configurations that come from different data splits are constituted of the same atoms. Test and validation data each consist of 5\% of the total data and are chosen such that configurations contain all heavy atoms (C, N, O). Potentially, models can learn fundamental features faster from simpler molecules, therefore, all molecules with less than three heavy atom types are kept in the training data. The models are trained on the training data with early stopping on the validation data, and we report the mean and standard deviation of \gls{rmse} and \gls{mae} from the evaluation of test data.

\subsubsection*{QM9 and QM9x}
QM9\cite{qm9} consists of \gls{dft} calculations of various properties for 135k small organic molecules in equilibrium configurations. All molecules in the dataset contain up to 9 heavy atoms, including C, N, O, and F. QM9 is ubiquitous as a benchmark for new \gls{qm} methods, and to enable direct comparison with Transition1x, all geometries from the QM9 dataset is recalculated with the appropriate level of \gls{dft}. Since configurations in the original QM9 \cite{qm9} are not relaxed in our potential, there will be forces on some configurations. All recalculated geometries are saved in a dataset that we shall refer to as QM9x. 

\subsubsection*{ANI1x}
ANI1x\cite{ani1x} is a dataset of off-equilibrium molecular configurations generated by perturbing equilibrium configurations using pseudo molecular dynamics. Data is included or rejected from the dataset based on the \gls{qbc} algorithm. In \gls{qbc} an ensemble (or committee) of models is trained on the dataset, and the relevance of new proposed data is assessed through the variance of the ensemble's predictions without having to perform expensive calculations on the data. It is assumed that data points will contribute new information to the dataset if the committee disagrees. It is cheaper to evaluate the committee on data than running \gls{dft} calculations, so it is possible to screen many candidate configurations before calculating force and energy with more expensive methods. The dataset is generated by alternating between training models and expanding the dataset. The procedure resulted in force and energy calculations for approximately 5 million configurations containing C, O, N, and H.

\section*{Data Records}[p]
\begin{figure*}
    \centering
    \includegraphics[scale=0.9]{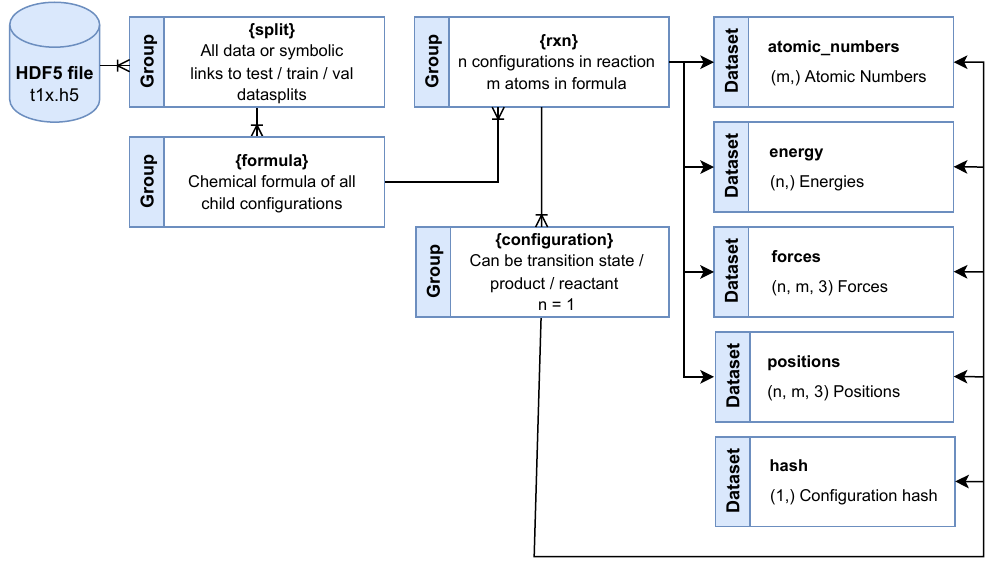}
    \caption{Structure of HDF5 file. Parent groups are data/train/val/test. The data group contains all configurations in the set, and the train/val/test groups contain the suggested data splits used in this paper. Each split has a set of chemical formulas unique to that split, and each formula contains all reactions with the given atoms. Finally, energy and force calculations can be accessed from the reaction groups for all intermediate configurations, including transition state, product, and reactant.}
    \label{fig:h5}
\end{figure*}

\begin{figure*}[p]
    \centering
    \includegraphics[scale=0.9]{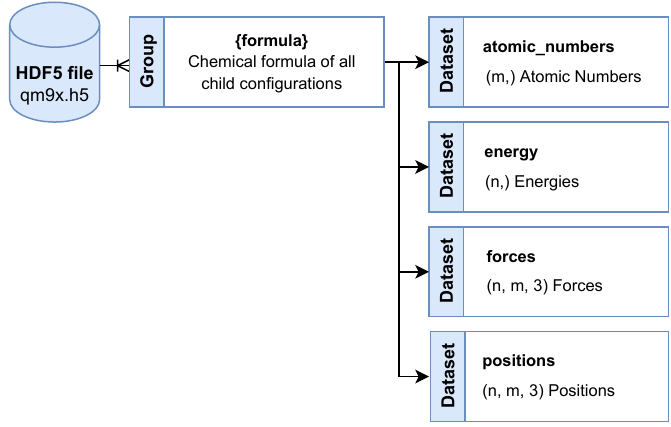}
    \caption{Structure of QM9x HDF5 file. Energy and force calculations for all configurations in the QM9x dataset consisting of a certain combination of atoms can be accessed as datasets through the formula group.}
    \label{fig:qm9x_h5}
\end{figure*}

Data records for Transition1x are available in a single \gls{hdf5}\cite{hdf5} file; Transition1x.h5, hosted by figshare \cite{Schreiner2022}. It can be downloaded at \href{https://doi.org/10.6084/m9.figshare.19614657}{here} or through the repository \href{https://gitlab.com/matschreiner/Transition1x}{https://gitlab.com/matschreiner/Transition1x}. The HDF5 file structure is as shown in Figure \ref{fig:h5}. The parent file has four groups, one group contains all data and the three other groups contain symbolic links to the train, test, and validation data -- these are the data splits used in this paper. In each data split group, there is a group for each chemical formula under which there is a subgroup for each reaction with the corresponding atoms. Each reaction has four datasets; the atomic numbers of the reaction, the energies of the configurations, the forces acting on the individual atoms, and the positions of atoms. For a reaction with $m$ atoms where we have saved $n$ images, the \emph{atomic\_numbers} dataset will have dimensions $(m,)$, one for each atom. The energy dataset will have dimensions $(n,)$, one energy per configuration. The force and position datasets will have dimensions $(n, m, 3)$ as we need three components of position and force for each of $m$ atoms in $n$ configurations. Under each reaction group, there is a child group for reactant, transition state, and product that follow the same structure as described above with $n=1$. Products from some reactions are reactants for the next, and they can be linked with a hash value available for each product, transition state, and reactant in the hash dataset. The data has been uploaded to figshare, and there is a git repository with data loaders that can turn the \gls{hdf5} file into an \gls{ase} database or save the configurations as .xyz files. 

Data records for QM9x are also available in a \gls{hdf5} file; QM9x.h5, hosted by figshare. It can be downloaded at \href{https://doi.org/10.6084/m9.figshare.20449701}{here} or through the repository \href{https://gitlab.com/matschreiner/QM9x}{https://gitlab.com/matschreiner/QM9x}. The HDF5 file structure is shown in Figure \ref{fig:qm9x_h5}. Energy and force calculations for all configurations in the QM9x dataset consisting
of a certain combination of atoms can be accessed as datasets through the formula group. The dimensions and structure of these datasets follow the same logic as described above.

\section*{Technical Validation}

\begin{figure*}[ht]
    \centering
    \includegraphics[width=\textwidth]{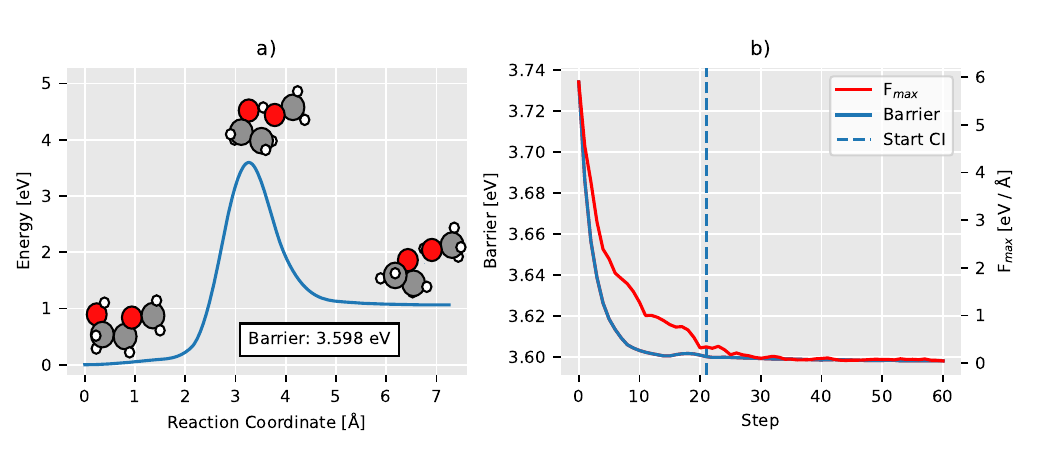}
    \caption{Plot of \gls{neb} convergence on example reaction. Panel a) displays the final \gls{mep} with reactant, transition state, and product plotted on top with H, C, and O in white, black, and red, respectively. On the x-axis; the reaction coordinate -- distance along the reaction path in configurational space, measured in Å. On the y-axis; the difference in potential energy between reactant and current configuration. Panel b) displays the convergence of \gls{neb}. On the x-axis; iterations of \gls{neb}. On the y-axis; force in eVÅ$^{-1}$ and energy barrier in eV at the current step. F$_{max}$, shown in red, is the maximal perpendicular force acting on any geometry along the path, and Barrier, shown in blue, is the height of the energy barrier found at the current step. Moving right in the plot both F$_{max}$ converges towards zero as \gls{neb} finds the saddle point, and the Barrier converges towards the final value of ~3.6 eV that can be seen in panel a. }
    \label{fig:converged}
\end{figure*}

\begin{figure*}[ht]
    \centering
    \includegraphics{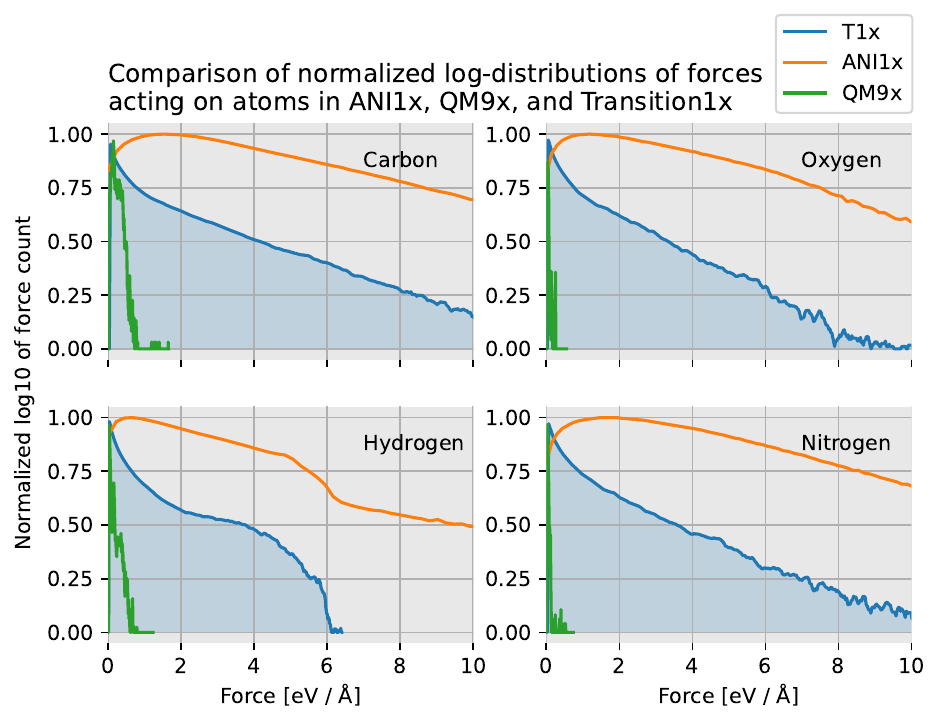}
    \caption{Distribution of forces acting on atom-types in each dataset. The x-axis is the force measured in eV/Å. The y-axis is the base 10 logarithm of the count of forces in each bin, normalized over the full domain so that all sets can be compared. In blue; Transition1x. In yellow; ANI1x. In green QM9x.}
    \label{fig:forces}
\end{figure*}

\begin{figure*}[p]
    \centering
    \includegraphics[width=\textwidth]{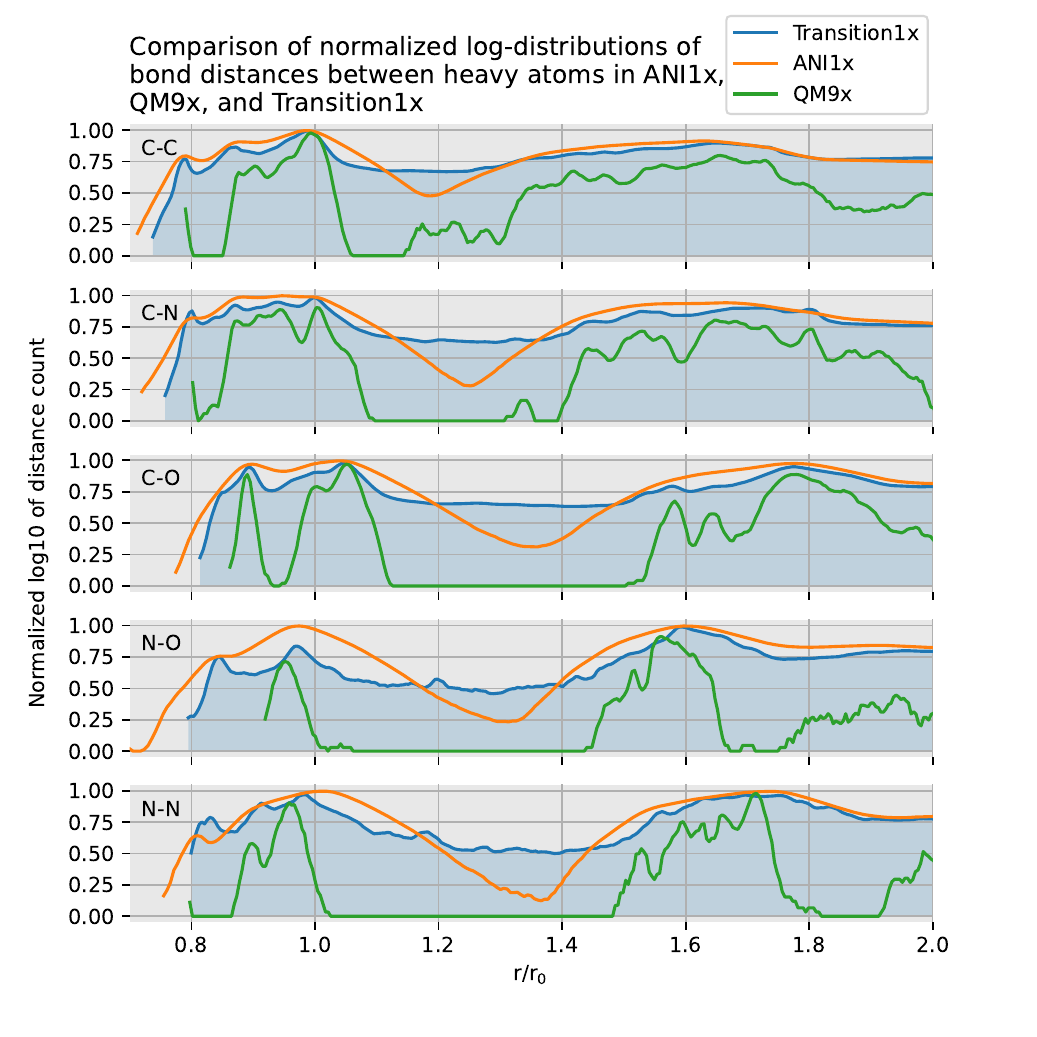}
    \caption{Distribution of interatomic distances between heavy atoms in each dataset. A configuration with n heavy atoms contributes with $n(n-1) / 2$ distances in the count. On the y-axis; the log frequency of interatomic distance, normalized between 0 and 1 for comparison as datasets vary in size. On the x-axis; distance given in units of $r_0$ where $r_0$ is the equilibrium bond length for a single bond between the smallest possible stable molecule that can be made with the atoms in question. In blue; Transition1x. In yellow; ANI1x. In green; QM9x, recalculated using our level of theory.}    
    \label{fig:dists}
\end{figure*}

\begin{table*}[ht]
\centering
\begin{tabular}{|ll|ll|ll|}
\hline
              &           & \multicolumn{2}{c|}{Energy {[}eV{]}} & \multicolumn{2}{c|}{Forces {[}eV/Å{]}} \\
              
Trained on    & Tested on & RMSE              & MAE              & RMSE               & MAE               \\
\hline

ANI1x         &  &0.629 (11)   &0.495 (10)  & 0.593 (18)   & 0.331 (7) \\
Transition1x           & Transition States & \textbf{0.112 (3)}  & \textbf{0.075 (1)} & \textbf{ 0.158 (1)}  & \textbf{0.089 (1)} \\
QM9x          &  & 3.132 (23)  & 2.957 (25) & 0.637 (15)  & 0.261 (4) \\ 

\hline
ANI1x         &  & \textbf{0.044(5)}   &\textbf{ 0.023(1)}  & \textbf{0.041(1)}   & \textbf{0.016(0)} \\
Transition1x           & ANI1x & 0.365(17)  & 0.226(8) & 0.321(25)  & 0.078(1) \\
QM9x           &  & 3.042(13)  & 2.313(11) & 1.314(8)   & 0.559(2) \\ 
\hline

ANI1x         &    & 0.628(63)  & 0.289(13) & 0.542(118) &0.13(5)  \\
Transition1x           &  Transition1x  & \textbf{0.102(2)}   & \textbf{0.048(1)}  & \textbf{0.102(1)}   & \textbf{0.046(1)} \\
QMx          &    & 2.613(18)  & 1.421(11) & 0.433(3)  & 0.199(1)\\ 
\hline

ANI1x         &    & 0.134(1)   & 0.124(1)  & 0.051(1)   & 0.025(1) \\
Transition1x           &  QM9x  & 0.111(2)   & 0.074(3)  & 0.072(1)   & 0.038(0) \\
QM9x           &    & \textbf{0.04(2)}    & \textbf{0.015(1) } &\textbf{ 0.014(0)}   &\textbf{ 0.005(0)} \\

\hline
\end{tabular}
\caption{Test results of PaiNN models trained on ANI1x, QM9x, Transition1x. We report RMSE and MAE on energy and forces. Force error is the component-wise error between the predicted and true force vector. The test sets have been constructed such that all configurations contain C, N, O, and H, and such that no formula has been seen previously in the training data. We show the best performing model in bold in each test-setup.}
\label{tab:eval}
\end{table*}

In figure \ref{fig:converged} we show the \gls{mep} for a reaction involving C3H7O2 and the convergence of \gls{neb} for it. Often the barrier grows initially after turning on the climbing image because we start maximizing the energy in a new degree of freedom. \gls{neb} converged on 10073 out of 11961 reactions, and 89 percent of the converged reactions did so within the first 200 iterations. To ensure the cleanliness of the data, we choose to discard all reactions that do not converge -- these reactions often contain unphysical structures that do more harm than good as training data. \\

The dataset includes a wide range of organic reactions. All reactions contain up to 7 heavy atoms including C, N, and O, and up to six bond changes where bonds are breaking and forming between all combinations of heavy atoms. Detailed analysis of the number of bond changes per reaction, number of bond changes involving specific pairs of atoms, the spread of activation energies, and SMARTS strings describing reactive centers of the reactions, is included in the original work Grambow et al. 2020 \cite{rxns}. \\

The perpendicular force drops off rapidly when running \gls{neb} and so does the variation in data between iterations as the path is nudged less between iterations towards convergence of the algorithm. $F_{max}$ is used as a proxy for how much the path moves between iterations and once the cumulative $F_{max}$ since the last included path exceeds a threshold of 0.1 eV/Å, the path is included in the dataset. \\

The transition states found with \gls{neb} correspond to the transition states from the original \gls{gsm} data with a MAE of 0.16 eV, RMSE of 0.19 eV, and an average Root Mean Square Deviation (RMSD) of 0.11Å. See Figure \ref{fig:compare} in the Appendix for details. Barrier energies match, but the \gls{neb} energies tend to be shifted higher. Generally, it is easier to describe electron clouds around relaxed configurations than around transition states where bonds are breaking and other complex interactions take place. Therefore, the more expressive basis set enables us to relax configurations around transition states further than around equilibrium states which results in lower barrier heights. There are more outliers above the $x=y$ line than below it which indicates that \gls{gsm} was caught in suboptimal reaction pathways more often than \gls{neb}.
\\

Figure \ref{fig:forces} displays the distribution of forces on each type of atom in Transition1x compared with ANI1x and QM9. Interestingly, even though geometries in Transition1x are further away from equilibrium than in ANI1x (regions between equilibria are actively sought out in Transition1x), the distribution of forces on ANI1x has flatter, wider tails signifying higher variance in forces. Moreover, Transition1x cusps at zero whereas ANI1x maxima lie further out. Large forces are not necessarily involved when dealing with reactive systems. When reaction pathways are minimized, forces are minimized too in all but one degree of freedom. The transition states contribute as much to the force distribution as equilibrium configurations, as the transition state is a saddle point with no net force on any atoms. ANI1x has no inherent bias towards low forces on geometries as it explores configurational space with pseudo-\gls{md} and therefore, even though the configurations are closer to equilibrium we see a higher variance in forces. On the heavy atoms, the tails are qualitatively equal between ANI1x and Transition1x, trailing off exponentially, but it is different for hydrogen. In ANI1x, forces on Hydrogen trail off exponentially as with the other atoms, but for Transition1x there is a sudden drop of the distribution.  Hydrogen atoms are often at the outskirts of the molecules and are relatively free to move compared to heavier atoms on the backbone. In the case of the Transition1x data generation procedure, energy and forces are minimized, and therefore Hydrogen atoms do not experience large forces as they have lots of freedom to relax in the geometry. In ANI1x, configurations are generated by perturbing the geometries randomly, and hydrogen atoms might end up with unrealistically large forces on them. This might be a general problem with ANI1x and also a reason why ANI1x is not a proper dataset to learn reaction mechanisms.
\\

Even though ANI1x has a wider distribution of forces, the inter-atomic distances between pairs of heavy atoms are less varied than in Transition1x. Figure \ref{fig:dists} displays the distribution of distances between pairs of heavy atoms for Transition1x, ANI1x, and QM9x. For QM9x, a dataset of only equilibrium configurations,  some inter-atomic distances are not present at all. Distances are measured in units of $r_0$, the single bond equilibrium distance between the atoms in the smallest possible molecule constructed out of the two. For example, in the case of "C, C" we measure in units of the distance between carbon atoms in ethane. Many of the more extreme inter-atomic distances in Transition1x are difficult to produce by the normal mode sampling technique of ANI1x as many atoms would randomly have to move such that the whole molecule moves along a low-energy valley. However, because \gls{neb} samples low-energy valleys by design, we discover likely molecules with inter-atomic distances that are otherwise energetically unfavorable.
\\

We test all resulting models against the test data from each dataset and Transition States from the test reactions. Table \ref{tab:eval} displays the results. It is clear from their evaluation of Transition1x and transition states, that models trained on ANI1x do not have sufficient data in transition state regions to properly learn the complex interactions present here. ANI1x has a broad variety of chemical structures, but many of the fundamental interactions found in ANI1x are present in Transition1x, which is why models trained on Transition1x perform better on ANI1x than vice versa. In general, the \gls{pes} of a set of atoms is an incredibly complex function of quantum mechanical nature. Models trained on QM9x do not perform well on either Transition1x or ANI1x. This is as expected as QM9x contains only equilibrium (or very close to equilibrium in the new potential) structures, so the models trained on it have not seen any of the out-of-equilibrium interactions that are present in the more challenging datasets of ANI1x and Transition1x.
\\ 

Transition state data is required if we want to replace \gls{dft} with cheap \gls{ml} potentials in algorithms such as \gls{neb}\cite{neuralneb} or \gls{gsm}, or train molecular dynamics models to work in transition state regions. \glspl{nn} are phenomenal function approximators, given sufficient training examples, but they do not extrapolate well. Training examples spanning the whole energy surface are needed to train reliable and general-purpose \gls{ml} models. Transition1x is a new type of dataset that explores different regions of chemical space than other popular datasets and it is highly relevant as it expands on the completeness of available data in the literature. 

\section*{Usage Notes}
There are examples and data loaders available in the repositories  \href{https://gitlab.com/matschreiner/Transition1x}{https://gitlab.com/matschreiner/Transition1x} and \href{https://gitlab.com/matschreiner/QM9x}{https://gitlab.com/matschreiner/QM9x}. There are scripts for downloading data in \gls{hdf5} format and easy-to-use data loaders that can loop through the files given only hdf5-paths. See the repositories and documentation for examples of how to use the data loaders and download the datasets.

\section*{Code availability}
All electronic structure calculations were computed with the ORCA electronic structure packages, version 5.0.2. All NEB calculations were computed with \gls{ase} version 3.22.1 The analysis scripts are available upon request. 

\section*{Author contributions statement}
M.S., A.B., T.V, and O.W. conceived the study. M.S. wrote the code, conducted the experiments, and wrote the majority of the article. A.B. and M.S. wrote background and summary, A.B. and O.W. provided supervision and reviewed the article, and J.B. reviewed the article and provided helpful discussions.

\section*{Competing interests}
The authors declare no competing interests.

\section*{Acknowledgements}
The authors acknowledge support from the Novo Nordisk Foundation (SURE, NNF19OC0057822) and the
European Union’s Horizon 2020 research and innovation program under Grant Agreement No. 957189
(BIG-MAP) and No. 957213 (BATTERY2030PLUS).
\\

\noindent Ole Winther also receives support from Novo Nordisk Foundation through the Center for Basic Machine Learning Research in Life Science (NNF20OC0062606) and the Pioneer Centre for AI, DNRF grant number P1.

\onecolumn
\bibliography{sample.bib}
\newpage

\appendix

\section{Comparison of GSM and NEB}

\begin{figure}[H]
\centering
\begin{subfigure}{.5\textwidth}
  \centering
  \includegraphics[]{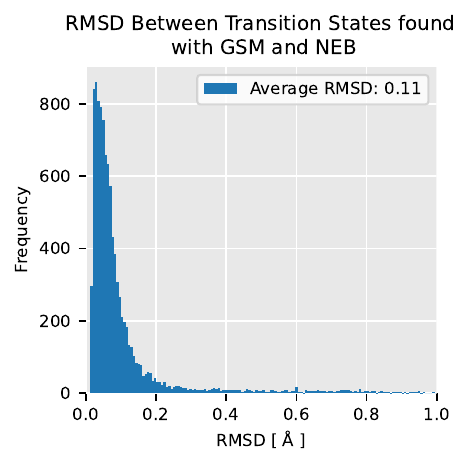}
  \caption{A subfigure}
\end{subfigure}%
\begin{subfigure}{.5\textwidth}
  \centering
  \includegraphics[]{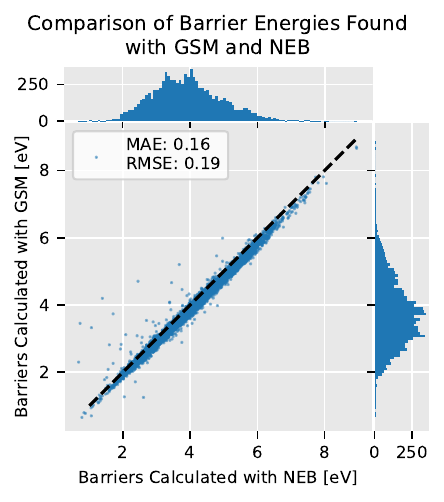}
  \caption{A subfigure}
\end{subfigure}
 \caption{Comparison of transition states and barriers found with \gls{neb} and the 6-31G(d) basis set in this work, and \gls{gsm} and the def2-mSVP basis set in the original work. Panel (a) displays a histogram of Root Mean Square Deviation (RMSD) between transition states found. Panel (b) displays energies in eV for all transition states calculated using \gls{neb} on the x-axis and \gls{gsm} on the y-axis.}
 \label{fig:compare}
\end{figure}

\end{document}